\documentclass[12pt,preprint]{aastex}
\shorttitle{Simulations of White Dwarf Disk Boundary Layers}
\shortauthors{Balsara et al.}
\begin{document}
\bibliographystyle{apj}

\title{Simulations of the Boundary Layer Between a White Dwarf
and its Accretion Disk}

\author{Dinshaw S. Balsara \& Jacob Lund Fisker}
\affil{Department of Physics, University of Notre Dame, Notre Dame, IN 46556 \\
dbalsara@nd.edu; jfisker@nd.edu}

\author{Patrick Godon\altaffilmark{1} \& Edward M. Sion}
\affil{Department of Astronomy and Astrophysics, 
Villanova University, Villanova, PA 19085 \\ 
patrick.godon@villanova.edu; edward.sion@villanova.edu}

\altaffiltext{1}
{Visiting at the Space Telescope Science Institute, Baltimore, MD 21218,
USA, godon@stsci.edu}

\begin{abstract}

Using a 2.5D time-dependent numerical code we recently developed, we solve the full compressible Navier-Stokes equations to determine the structure of the boundary layer between the white dwarf and the accretion disk in non-magnetic cataclysmic variable systems.  In this preliminary work, our numerical approach does not include radiation. In the energy equation, we either take the dissipation function ($\Phi$) into account or we assume that the energy dissipated by viscous processes is instantly radiated away ($\Phi=0$).  For a slowly rotating non-magnetized accreting white dwarf, the accretion disk extends all the way to the stellar surface.  There, the matter impacts and spreads towards the poles as new matter continuously piles up behind it.  We carry out numerical simulations for different values of the alpha viscosity parameter ($\alpha$), corresponding to different mass accretion rates.  In the high viscosity cases ($\alpha = 0.1$), the spreading boundary layer sets off a gravity wave in the surface matter. The accretion flow moves supersonically over the cusp making it susceptible to the rapid development of gravity wave and/or Kelvin-Helmholtz shearing instabilities.  This BL is optically thick and extends more than 30 degrees to either side of the disk plane after only 3/4 of a Keplerian rotation period ($t_K$=19s).  In the low viscosity cases ($\alpha =0.001$) , the spreading boundary layer does not set off gravity waves and it is optically thin.

\end{abstract}


\section{Introduction: Accreting White Dwarfs in Cataclysmic
Variables}\label{sec:introduction}
   
Cataclysmic variables (CVs) form an interesting class of short-period
close binary systems comprising a hot white dwarf (WD) and a relatively
lower mass red dwarf star filling its Roche lobe \citep{cra56,kra62}.
In such systems hydrogen-rich matter from the red dwarf exits through
the inner Lagrange point (L1) and flows towards the white dwarf.  
In the absence of strong magnetic fields, the
matter forms an accretion disk around the white dwarf due
to the excess angular momentum originating from the orbital motion of
the binary \citep{Prendergast68,fla74,lub75}. On-going accretion at a low rate
(quiescence) is interrupted every few weeks to months by intense
accretion (outburst) 
of days to weeks - a dwarf nova (DN) accretion
event \citep{bat72}, thereby increasing the luminosity of the systems by several
magnitudes.  A thermal instability in the accretion disk is believed to
trigger the increase in the mass transfer rate ($\dot{m}$)  
through the disk and thus an increase in the rate of gravitational
energy release \citep{Cannizzo88}.

CV systems are divided in sub-classes according to the duration,
occurrence and amplitude of their outburst \citep{hac93,Warner95,rit98}:   
e.g. dwarf nova systems (DNs) 
are non-magnetic and accrete through a disk, they spend most
of their time in the quiescent state; nova-like systems (NLs) are
disk-systems found mostly in the high outburst state; polars are
devoid of an accretion disk due to their strong magnetic fields (the
matter is funneled through the magnetic field lines onto the poles
of the WD); and intermediate polars (IPs) have truncated inner
disks due to their moderately strong magnetic fields. 
Another CV subclass is the classical nova (or just ``nova'';), 
characterized by an episode of unstable thermonuclear burning
(the thermonuclear runaway - TNR; \citet{ros68}).
All CV system are believed to undergo a classical nova explosion
every few thousand years or more, when enough hydrogen-rich
material accumulated in the envelope to reach the ignition point
at the electron-degenerate base of the envelope - where it is  
compressed under the large gravity of the WD 
\citep{sta71a,sta71b,Starrfield72}.
In the present work we concentrate on the study of the non (or
weakly) magnetized DN systems, where the accretion disk extends
all the way to the surface of the WD. 

Since the white dwarf is the most common end-product of stellar evolution
($\approx 90$\% of all the stars in the Galaxy have evolved or will
evolve into white dwarfs) and the accretion disk is the most
common universal structure resulting from mass transfer with angular 
momentum, and both can be directly observed in cataclysmic variable
systems (in the ultraviolet), an understanding of the accretion process
in cataclysmic variables is the first step in a global understanding
of accretion in other systems throughout the universe. These include
young stellar objects, accretion onto neutron stars and black holes,
and the most difficult to study, active galactic nuclei.
Therefore, accreting white dwarfs in cataclysmic variables are the best
{\it astronomical laboratories} to study the physics of accretion
disks. \\  

\subsection{The Accretion Disk \& Boundary Layer in One-Dimension}  

In the disk, magneto-hydrodynamic (MHD)
turbulence \citep{Shakura73} due to a magneto-rotational
instability \citep{Balbus94a} dissipates potential energy and transfers  
angular momentum outward. As a result, the disk matter slowly spirals
inwards towards the white dwarf \citep{Pringle81}.
The total potential energy of accretion can be released at the
maximum rate 
\begin{equation} 
L_{acc}=\frac{GM_*\dot{m}}{r_*} = \dot{m} r_*^2 \Omega_K^2(r_*) , 
\end{equation} 
where $G$ is the gravitational constant, $M_*$ is the mass of the
WD, $r_*$ its radius, $\dot{m}$ is the mass accretion rate, and  
$\Omega_K(r_*)$ is the Keplerian angular velocity at one 
stellar radius $r_*$.
This is the maximum amount of energy that can be extracted from the
accretion process per unit of time 
and the actual luminosity can be smaller than this
(see below). 
In the standard disk theory \citep{Shakura73,Lynden-Bell74},
the accretion disk is axisymmetric, geometrically thin in the
vertical dimension (it has a thickness $H$ such that $H/r<<1$),  
and the energy dissipated by the (turbulent)
viscosity is instantly radiated locally in the $\pm z$ directions.  
Only half of the accretion luminosity is emitted by the disk
($L_{disk}=L_{acc}/2$), since the matter is still moving at a nearly
Keplerian velocity, $v_K\approx \sqrt{GM_*/r_*}$, before it is ultimately
accreted \citep{Lynden-Bell74}.

The remaining accretion energy must, therefore, be dissipated
in order for the matter to be accreted onto the surface
of the more slowly rotating WD \citep{Pringle81}.
The disk matter is decelerated in that region where the inner disk
reaches the stellar surface: the boundary layer (BL).  
The height of the BL can be comparable to 
the scale height $H$ of the accretion disk.
The remaining rotational kinetic energy of the accreting matter 
dissipated in the
boundary layer per unit of time ($L_{BL}$) is nearly equal to half of the
total luminosity of the accreting matter ($L_{acc}$),
namely:
\begin{equation}  
L_{BL}=
\left( 1 - \beta^2 \right) L_{disk} =  
\frac{\dot{m} r_*^2}{2} ( \Omega_K^2(r_*) - \Omega_*^2 ) , 
\end{equation} 
where $\beta=\Omega_*/\Omega_K(r_*)$,  
and $\Omega_*$ is the stellar angular velocity.
This is so because the material accreted onto
the WD surface corotates with it and keeps
a fraction of the rotational kinetic energy.  
\citet{Kluzniak87}, however, pointed out that  
part of the BL energy goes into
spinning up the accreting star through the shear
at the stellar equator, and hence, the fraction
that can be radiated away is smaller than the one given in eq.(2). 
Namely, one has  
\begin{equation}  
L_{BL}=
( 1- \beta )^2  L_{disk} =  
\frac{\dot{m} r_*^2}{2} ( \Omega_K(r_*) - \Omega_* )^2.  
\end{equation} 
Therefore, for a non-rotating WD: $L_{BL}=L_{disk}=L_{acc}/2$.
However, for a rotating WD $L_{BL} < L_{disk}$,
and the total energy
radiated from the accretion process ($L_{BL}+L_{disk}$)  
is actually smaller than $L_{acc}$ as defined by eq.(1).   
The energy ``kept'' by the corotating accreted material 
per unit of time is :  
\begin{equation} 
\beta^2 L_{disk} =
\frac{\dot{m} r_*^2}{2} \Omega_*^2,  
\end{equation} 
and the power invested to spin up the star through the shear at the equator
(eq.2-eq.3) is:
\begin{equation}
[ (1-\beta^2) - (1-\beta)^2 ] L_{disk} =  
2 \beta ( 1- \beta  )  
L_{disk}  = \beta ( 1 - \beta ) L_{acc} . 
\end{equation}   
The above relations are correct to order of
$\eta = 1 - r_m/r_*$, where $r_m$ is the radius at which the gradient
of the angular velocity changes sign $\partial \Omega / \partial r =0$
(for further details of the one-dimensional analysis
see also \citet{Regev83,Kley91,Popham95}).
Usually $\eta$ is of the order of $H/r$ or smaller.
For a WD star rotating at 10\% of the Keplerian velocity 
(i.e. several 100km/s)
the BL is expected to radiate $L_{BL}=0.81L_{disk}$, while a fraction
of $0.18L_{disk}$ is invested in spinning up the star.
For a WD star rotating at 30\% of the Keplerian velocity 
(i.e. $\sim 1000$km/s)
the BL is expected to radiate $L_{BL} = 0.49L_{disk}$. 
For a star rotating near break-up the angular velocity in the disk does not
have an extremum and decreases steadily outward, therefore 
the disk is expected to spin-down the fast rotating WD
\citep{pop91}. In this case there is no boundary layer.  

Since the BL is much smaller than the disk itself and each radiate
almost equal amounts of energy, the disk emits mainly in the optical
and near UV, whereas the much hotter BL emits in the far UV and in X-rays.
During outburst, when $\dot{m}\approx 10^{-9}-10^{-8} M_{\odot}$/yr
\citep{war87,Cannizzo88}, observations (e.g. \citet{cor80,mau95,mau04}) 
reveal that the BL is optically thick 
and emits mainly in the FUV and soft X-ray bands 
with an effective temperature $T_{eff}$ of a few $10^5$K
as expected by the theory 
(\citet{Pringle79,Regev83,Godon95,Popham95,Collins98b,Obach99}).  
During quiescence, when $\dot{m}\approx 10^{-12}-10^{-10} M_{\odot}$/yr
\citep{war87}, observations (e.g. \citet{muk04,Pandel03,Pandel05}) 
reveal that the boundary layer is optically thin 
and emits in the hard X-ray band with a temperature
of the order of $10^8$K, as expected by the theory 
(e.g. \citet{Pringle79,Tylenda81,King84,Shaviv87,Narayan93,Popham99}).  
However, in addition, in the outer region of the BL, where the BL
meets the disk, the optical thickness becomes larger again ($\sim 1$)
and that region emits in the FUV with a temperature $T_{eff}$ reaching
$\sim 10^5$K \citep{Popham99}.

\subsection{Observational Background}  

Observationally, it was shown almost three decades ago 
that
non-magnetic CVs do emit some fraction of their luminosity in the
X-ray bands (e.g. \citet{cor80,Cordova81,Becker81,Cordova83,Patterson85};
the ROSAT All-Sky Survey \citep{Beuermann93}).
As expected from the one-dimensional standard disk and boundary layer
theories, during quiescence hard X-rays (10keV and
higher) were observed from a small region close to the WD (e.g.
\citet{vanTeeseling96,Mukai97}), while during
outburst this emission is replaced by soft X-ray and EUV
(see e.g. the review of \citet{Mauche96} and the references therein).
The EUV region of the spectrum
is very difficult to observe because the absorption cross section
of (ISM) neutral hydrogen is very high. Because of this, only a
few systems have been successfully observed in the EUV, such as e.g. VW Hyi
(\citet{Mauche91}; for which $N(H)\sim 6\times 10^{17}$cm$^{-2}$,
\citet{Polidan90}).
Assuming that the X-ray and EUV emissions are from the BL, and
the optical and UV emissions are from the disk, many previous studies
found a very low ratio $L_{BL}/L_{disk}$ ($\approx 0.001-0.04$,
\citet{Mauche91,Hoare91}) during outburst, and a ratio of
$L_{BL}/L_{disk}\approx 0.25$ in quiescence (e.g. for VW Hyi,
\citet{Belloni91} assuming that the WD contributes 50 percent 
of the UV flux). 
The X-ray observations of {\it underluminous} boundary layers culminated  
with the {\it ROSAT} observations of ten cataclysmic variables 
(BA Cam, YZ Cnc, GP Com, VW Hyi, WX Hyi, TY PsA, V3885 Sgr, 
CY UMa, CY Vel, IX Vel)
by \citet{vanTeeseling94}, in which 8 systems 
were caught in quiescence
and 2 systems were caught in outburst. \citet{vanTeeseling94} 
derived X-ray fluxes from their observations and UV fluxes from
existing {\it IUE} observations and found that ratio of the X-ray Luminosity 
to the UV+Optical Luminosity is much smaller than one.  

Many processes were discussed to explain the "missing
boundary layer" (e.g. \citet{Ferland82,Shaviv87,King97,Meyer94,Ponman95}).  
The main idea was that the kinetic energy of the BL could also be
converted into winds (e.g. \citet{King84}), WD or belt rotation
(e.g. \citet{Kippenhahn78}),  
heating (e.g. \citet{Shavivand87,Regev89};  
or maybe advected into the outer stellar envelop
(e.g.. \citet{Godon96,Godon97,Popham97}). 
Some systems have been observed to have a WD rotating at a rather large
rotational velocity, of the order of $\sim 1,000$km/sec
(e.g. \citet{che97,Pandel05}), which implies (from eq.3) 
$L_{BL}/L_{disk} = 1/2$. 
However, the observed X-ray luminosities have been much smaller than
this and would imply a near-Keplerian rotation rate for many systems,
which  is not the case.  

Recent XMM-Newton observations \citep{Pandel05},
taking into account the contribution
of the disk, WD and BL in the optical, UV and X-ray bands, found no
evidence of an underluminous BL in 8 quiescent dwarf novae 
(OY Car, WW Cet, AB Dra, U Gem, VW Hyi, T Leo, TY PsA, SU UMa) 
and the
data are consistent with $L_{BL}\approx L_{disk}$.
The main difference with previous studies was that the WD contribution
was taken into account with realistic temperatures taken from the literature
and could dominate over the disk in the UV.
\citet{Pandel05} basically assumed $L_{disk}=L_{UV}+L_{opt}-L_{WD}$
and $L_{BL}=L_X$, and found $L_{BL}/L_{disk} \sim 1$ for 6 objects
among 8. For VW Hyi and U Gem (with $L_{BL}/L_{disk} \sim 1/3, 1/2$ 
respectively) they suggest that the inner disk is
truncated at $r\sim 3R_{WD}$ to explain the discrepancy.  

\citet{Godon05} further computed the contribution of 
all the emitting components (WD, disk, BL) in the optical, UV and 
X-ray based on the existing multiwavelength 
observations of VW Hyi, the standard
disk model \citep{Pringle81} 
and \citet{Popham99}'s model of the boundary layer. 
\citet{Popham99} has shown that (in quiescence) the BL emits part 
of it energy in the UV band from that region where 
the outer edge of the BL meets (and radiates energy into)  
the optically thick inner edge of the disk.  
\citet{Godon05} suggested that the
second component often observed in the quiescent UV spectra of DNe
(the so-called accretion belt also detected in the {\it FUSE} spectra of
VW Hyi --- \citep{god04}) is the UV emission from the outer BL.  
Taking the UV contribution of this second component 
($L_{2nd}$) into account \citet{Godon05} showed that
the luminosity of the BL of VW Hyi in quiescence 
($L_{BL} = L_X + L_{2nd}$) is as expected
from the theory (namely $L_{BL}=L_{disk} = L_{opt} + L_{UV} - L_{WD}$), 
therefore agreeing with \citet{Pandel05} that 
{\it there is no missing boundary layer} even for VW Hyi and without
disk truncation. 

On the other hand, spectroscopic UV observations
of accreting WDs have also reached a rather advanced stage.
Observations have been carried out for a range of CVs, 
\citep[to cite just a few]{Long93,Froning01,Szkody02,Welsh03,Sion04c,Godon04}
in quiescence and outburst. The
observations suggest that the spectrum is made up of several parts,
specifically, the underlying WD, the accretion disk, the accretion belt
that might form on the surface of the star, the hot spot
(where the matter inflowing from the L1 point hits the outer
rim of the accretion disk).
The temperature of each of these components proves to be a most important
diagnostic, though as better data come in, we can hope that other
hydrodynamical variables also become important diagnostics. The case of
VW Hydri is especially important because STIS measurements \citep{Sion04c} 
have captured that object during its transition to outburst. 
The VW Hyi transition to outburst caught by {\it STIS} was apparently
an outside-in outburst. The UV lagged the  optical. The
{\it STIS} showed that longer FUV wavelengths changed much faster
and manifested the flux sooner than shorter wavelengths.  
It has been found that the temperature of the accreting star can be
raised by as much as 50\% during outburst. The elevated WD temperature
returns to its quiescence value soon after the outburst.
The elevated surface temperature of the star may, therefore, be a good
diagnostic of the energy dissipated as the accretion stream impacts,
spreads out and slows down and reaches co-rotation with the star. 
The rate at which the WD cools following the heating during the 
outburst is not only diagnostic of the mass/depth of the heated surface
layers but also potentially a good diagnostic of the  optical depth
of the accreted material. 
A detailed understanding of BL physics may
also help one understand the so-called UV delay where the rise in the
UV emission lags the rise in the optical by several hours, indicating
that the UV emission might be more symptomatic of the physics of the
BL \citep{Livio92}. 
It is important to note here that 
our present numerical simulations  are relevant 
to the changes in the WD temperature observed on 
a time scale of the order of outburst/quiescence cycle (e.g. VW Hyi), 
and do not apply to compressional heating taking place 
on a much longer evolutionary time scale 
\citep{Sion95,Godon02,Godon03,Piro05}.  
\\  

\subsection{The Two-Dimensional Boundary Layer}

The BL has typically been solved in a model where the averaging of
the vertical structure which is employed in accretion disk studies
\citep{Shakura73} has been extrapolated to the surface. One also makes
the additional simplifying assumption that the vertical velocity in
the boundary layer is zero, just as it is in the outer parts of a
thin disk model. This reduces the description of the BL to a one
dimensional problem in the radial direction \citep{Pringle81,Meyer82,
Popham95,Collins98b}, where one actually solves the ``slim disk''
equations for the BL region (namely, the one-dimensional
disk equations + radiation in the radial direction; as first
proposed in the pioneering work of \citet{Regev83}).  
However, this assumption is not necessarily correct since the BL may
spread out  \citep{Ferland82}. The best one-dimensional models of
the BL 
\citep{Popham95} predict a rise in the BL temperature during outburst
that is much larger than the observed one.

Recognizing the multi-dimensionality of the problem several authors
attacked the problem directly using numerical methods
(e.g. \citet{Robertson86})
culminating in simulations which included flux-limited radiative
transport as well as different viscosities \citep{Kley87a,Kley89a,Kley91}.
However, while the efforts of \citet{Kley87a,Kley89a,Kley91} were
very bold and impressive, the high numerical resolution needed to
solve the problem has until now precluded following 
the problem numerically
on a long evolution time scale and or resolving the fine structure of
the flow.  
The importance of the meridional flow was demonstrated 
in protostars and FU Ori stars already by \citet{Kley96,Kley99}, who
show how matter spread to the poles to completely engulf the accreting
star. However,
in accretion onto a compact star the problem has not been  
solved because of the much smaller scales.

As the numerical two-dimensional simulations failed to follow the
evolution of the accretion onto the surface of the compact star,
an analytical approach was developed by  
\citet{Inogamov99},
who implemented the one-dimensional BL equations with
an analytic treatment of the meridional direction (assuming a shallow-water
equation), where the vertical
velocity in the boundary layer is permitted to be non-zero.  
This analytical treatment was carried out for an
accreting neutron-star (NS), and it was shown that the BL could
``spread'' and cover a significant part of the NS. 
Later on, \citep{Piro04} applied Inogamov \& Sunyaev's treatment 
of the ``spread layer'' to accreting white dwarf in outburst (the optically
thick case).  
The analytical work of \cite{Piro04} indicates that the  
numerical setup must include sufficient resolution in the radial and
meridional planes to capture the dynamics and fine structure of the BL.
It is the purpose of the present simulations to actually provide such
a numerical study, by following the evolution of the accretion onto
(and into!)
the surface of the WD in the inner region of the disk and close to
the equatorial plane ($i<30$deg).  

In the one dimensional picture the energy kept by the corotating
accreted material per unit time is given by eq.4. In two dimension,
however, as the matter possibly spreads evenly onto the WD surface,
its moment of inertia is rather similar to that of a spherical shell
and the energy kept by the accreted material per unit of time becomes
\begin{equation}
\frac{1}{2}\Omega_*^2 \frac{2}{3} \dot{m} r_*^2 = \frac{2}{3} \beta^2 L_{disk},
\end{equation} 
and the remaining energy (eq.4)-(eq.6)  
\begin{equation}
\frac{1}{3} \beta^2 L_{disk}
\end{equation} 
is available to further spin up the WD. 
As the matter moves toward the poles, to spread evenly on the WD surface, 
it has excess of angular velocity/momentum (due to the
differential velocity) which is added to the accreting WD.  
We therefore see that even for the most simplistic two-dimensional
model of the boundary layer, the spreading of matter on the WD
surface involves a significant fraction of the accretion energy 
(e.g. up to 10\% of the disk luminosity for a star rotating at
1,000km/sec --- eq.7).  
In the case of a star rotating near break-up velocity, accretion
at the equator {\it adds angular momentum to the star as the matter
spreads toward higher latitudes}. This contributes to a balancing
effect to the spin-down of the an accreting star rotating near
break-up as described by \citet{pop91}.

\subsection{The Importance of the Boundary Layer}

The structure of the BL is important to help disentangle 
and physically characterize
the different emitting components in an accreting WD system. 
The details of the emitted spectrum
depends sensitively on the detailed structure of the boundary 
layer and how it changes in response
to enhanced rates of accretion \citep{Fisker05c}. The boundary
layer is also important because it may play a role in the
still uncertain mechanism which drives the outflowing
bi-polar winds seen in dwarf novae during outburst and in nova-like
variables during their high optical brightness states. 
Our boundary 
layer simulations will eventually provide  the theoretical framework 
required to physically interpet the X-ray, EUV and FUV emission 
observed in compact binaries containing accreting degenerate stars.

However, the structure of the BL is also important because it
determines how the accreted matter ultimately distributes itself in
the envelope of the WD, which is important for classical novae.
In classical novae,
the thermonuclear runaway (TNR) ejects part or all of
the envelope. The initial CNO composition of the burning material
should be strongly enhanced compared to the accreted material to
account for composition of the observed ejecta \citep{Starrfield72}.
Several mechanisms have been suggested for this CNO enhancement (see
\cite{Jose05} and references therein). They can be roughly divided
into pre-burst mixing between accreted material and the underlying CO
rich WD \citep{Kippenhahn78,MacDonald83,Rosner01,Alexakis04} by accretion
driving instabilities or mixing with the underlying material when the
convective zone of the thermonuclear runaway extends deep enough to
dredge up CO material \citep{Starrfield72,Glasner97}.

However, before an exploration of the radiation emission characteristics
of the boundary layer can be carried out with full radiation hydrodynamics, 
we must understand the dynamical processes that lead to the formation 
of the boundary layer.
In this first stage our goal is to calculate the dynamical structure of the boundary layer with sufficient numerical resolution to capture the dynamical
evolution of the accretion flow and its interaction with the stellar
surface (see \cite{Fisker05c}), and we leave the
radiation-hydrodynamical study for later.
 
Our model is presented in section~\ref{sec:model}, the results are given
and discussed in section~\ref{sec:results}. Section~\ref{sec:conclusions} 
presents the conclusions. The equations we are solving are written
down explicitly in Appendix~\ref{appendix:model} 
while the initial and boundary conditions are described in 
Appendix~\ref{appendix:domainsetup}.

\section{Computational model}\label{sec:model}

The source of the angular momentum transport in the BL probably
involves a combination of magnetic fields and turbulence. Since the
angular velocity in the BL is not a decreasing function of radial distance
from the star, it is not clear that the magnetorotational instability (MRI)
\citep{Balbus94a} provides angular momentum transport in the BL. 
Actually, \citet{Brandenburg96} found that the effective alpha viscosity
parameter $\alpha$ would go to zero for a rotation law
$\Omega \sim r^{-q}$ when $q<0$. Nor is the
MRI essential at the BL because the accretion disk can directly exchange
angular momentum and mass with the outer layers of the star if an efficient
coupling mechanism is found between the disk and the star.  The source of
viscosity at the BL is still debated in the literature (see
\cite{Popham95,Piro04,Godon05}).
\citet{Inogamov99} assumed that the viscosity is due to purely
hydrodynamical turbulence as in the deceleration of subsonic or supersonic
flow above a solid surface. The case for a purely hydrodynamical
turbulence in the boundary layer was already debated earlier
\citep{Zahn90,Dubrulle93},
however the physical process is likely to take place
in three dimensions \citep{Orszag80}, e.g. by means of
streamwise vortices (e.g. \citet{Hamilton94}).
The most common instability in a flow over a {\it curved surface},
is that of the boundary layer flow over a concave surface unstable to the
centrifugal instability (as it violates Rayleigh's criterion for stability
--- \citep{Rayleigh16}). This instability leads to turbulence through the
formation of the G\"ortler vortices
\citep{Saric94}, which tap energy from the laminar flow and poor it into
the turbulence (this is a non-linear instability leading to a subcritical
transition to turbulence).
However, the boundary layer flow on a convex surface is not subject to this
instability, rather the opposite, the centrifugal force in such a
flow is "restoring".  Therefore, the instability that is the most likely
to take place in the star-disk boundary layer is the Kelvin-Helmhotz
shearing instability which will occur for a sufficiently large shear
(Richardson criteria, e.g. \citet{Drazin81}).

Following
\citep{Shakura73} , we parametrize the efficiency of the angular momentum
transport with an $\alpha$ viscosity prescription. In a multidimensional 
calculation $\alpha$ has to be spatially concentrated at the accretion disk 
and the stellar surface and the formulations developed in \citep{Papaloizou86} 
and \citep{Kley91} are used here. Specifically, we 
used $\nu = \alpha c_s min ( H, h_p)$ where $H$ is the scale height of the disk
and $h_p$ is the local pressure scale height in the boundary layer.
For these calculations, the pressure scale height in the boundary layer
that develops on the surface of the star is always smaller than the
scale height of the disk. The compressible Navier-Stokes
equations themselves are written explicitly 
in the Appendix \ref{appendix:model}.
Describing the angular momentum transport with a simple shear coefficient
means that the dynamics follows the Navier-Stokes equations.
Here the Navier-Stokes equations as given by \cite{Mihalas84} are solved
in spherical coordinates ($r$,$\theta$,$\phi$) on an axisymmetric mesh
with 384 ratioed zones in the radial direction and 128 ratioed zones
in the meridional range spanning 0 to 30 degrees from the disk plane
-- same as \cite{Fisker05c}.
The star was taken to be a non-rotating $0.6M_\odot$ WD with a radius
of $9\times 10^{8} cm$ . The radial extent of our computational domain
extended from $8.9\times 10^{8} cm$ to $1.1\times 10^{9} cm$. The
r-directional zones were concentrated towards the inner radial boundary
with each zone being 1\% larger than the previous one. The $\theta$-directional
zones were also ratioed with the smallest zones being closest to the
equator and each zone being 1.9\% larger than the previous one.
Such a ratioed zoning makes it possible for us to capture five scale
heights of the star's atmosphere as well as the vertical structure of
the disk. The zone ratioing also concentrates zones around the disk-star
interface. Unlike \citep{Kley91} , who used a mesh with 85  
zones in each of the radial and meridional directions, our mesh has
substantially better resolution. The scale heights of WD atmospheres
are now known to be substantially smaller than estimated in
\citep{Kley91,Kley87a}, making the larger meshes used in this work more
essential. The higher resolutions, as well as the use of higher order
Godunov methods, enable us to substantially reduce the role of numerical
diffusion.

In this paper, we wish to study not just the spin-down of the accretion
disk but also the spin-up of the star. For that reason, we retained five
scale heights of the WD's outer atmosphere. Care was taken to resolve
each stellar scale height in the radial direction with at least
twenty zones, which ensures a
numerically accurate, well-resolved and stable stellar atmosphere.
To ensure good resolution of the disk's structure, we retained at
least fifty zones across a disk scale height in the meridional direction.
The physical conditions 
describing the structure of the disk are given in Appendix 
\ref{appendix:domainsetup}.
Typical surface temperatures  for hydrodynamically accreting WDs are
$\sim$ 30,000K and typical disk temperatures are usually taken to
be $\sim$ 100,000K. This choice of temperatures would make the scale heights
of the disk and WD atmosphere too small to be resolved with the above-mentioned
resolutions. For this reason, we systematically allow the temperatures
of both the disk and WD atmosphere to be one order of magnitude larger
than the physical values. As a result, the simulations were carried
out with a stellar temperature of T=300,000K
and a disk temperature of T=1,000,000K. While this might seem be a hot
temperature for a disk in a CV, it is still much less than the virial
temperature and the corresponding disk thickness is $H/r=0.03$
instead of $H/r=0.01$ for a $10^5$K disk. The sound speed used in the
formulae for the viscosity $\nu$ was derived for the temperatures used
in the simulations. 
We assume a very tenuous and very hot halo which consists of the same material
as the disk (and the star). The mass in this halo is extremely small, making the
halo dynamically unimportant and its
sole use is to provide pressure balance at the upper
boundaries of the accretion disk
and the star. The fluids that make up the disk, halo and stellar atmosphere were
tagged with passive scalars and we are, therefore, in a position to assert that
the halo gas simply provides pressure support with minimal mixing into the
gas that makes up the disk or the stellar atmosphere.
Such a model with a disk and halo that are in dynamical equilibrium
with each other was briefly described in \cite{Balsara04}. 
Since that previous work did
not include the stellar atmosphere, appendix 
\ref{appendix:domainsetup} describes
the physical conditions that were assumed to set up 
the WD atmosphere, the accretion disk and the halo.
Symmetry across the disk plane is assumed.

Observations, and their interpretation in the context of the
\cite{Popham95} model have suggested that the interesting values    
of $\alpha$ range from 0.1 during outburst to 0.001 during quiescence.
\citet{Godon05} derived $\alpha=0.004$ in the boundary layer from
the XMM-Newton observations of VW Hyi in quiescence \citep{Pandel03}.  
For that reason, we performed
a parameter study with $\alpha=0.1$, $\alpha=0.03$, $\alpha=0.01$,
$\alpha=0.005$, and $\alpha=0.001$ . The same value of $\alpha$ was used
in the boundary layer and in the disk. This would seem to be a very reasonable
assumption if the same physical mechanism (such as a thermal instability,
\cite{Cannizzo98} or magnetic instability \cite{Livio98}) were to make
the disk fluid and the boundary layer fluid turbulent. Our simulations
do not include the effect
of radiative transfer in the boundary layer. It is worth noting that
in the optically thick limit, the BL retains most of the energy that is
generated by the viscous stresses. As a result, by retaining the viscous
energy generation terms in 
eq.~[\ref{eq:heatequation}] and assuming that the alpha viscosity parameter
is large ($\alpha \sim 0.1$) we mimic the situation where the
boundary layer is optically thick during outburst. 
Simulations were also carried out
by excluding the viscous energy generation terms in 
eq.~[\ref{eq:heatequation}] and assuming  $\alpha \sim 0.001$; in
that situation we mimic the limit where the boundary layer is 
optically thin during quiescence, i.e. the
BL radiates away all the heat that is generated by the viscous stresses.
Our results, therefore, bracket 
the two extreme cases. Should the high $\alpha$
simulations produce BLs with optical depths in excess of $50$ or $100$ , the
optically thick runs would find direct applicability to the astrophysical
problem. Likewise, should the low $\alpha$ simulations produce optically thin
boundary layers, the optically thin runs would find direct applicability to the astrophysical
problem. We show in the course of this 
work that such trends are indeed observed in
the simulations. In future work, we will include a treatment of the radiative
transfer terms in the BL thus obtaining results that are free of current limitations.
The full range of simulations that we report on here with the various values of
$\alpha$ and the inclusion or exclusion of viscous energy generation terms are
listed  in Table 1.

Since the inner radial boundary extends into the star,
we enforced reflective boundary conditions at that boundary. While such
a boundary condition would reflect any waves that reached
the star's surface, we find that in practice the waves do not propagate
to this depth in the present simulations. The use of such a reflective
boundary condition represents a compromise. While it might reflect back
waves that propagate more than five scale heights into the star, our
experience has shown that surface waves almost never propagate to that
depth. The present boundary conditions do have the positive
consequence that they prevent any unexpected mass or momentum 
inflow into the computational domain from the rest
of the star. 
The outer radial (computational) boundary was designed to respond to
inflow or outflow of fluid at the outer open boundary, and it 
was therefore treated by imposing the boundary conditions on the 
characteristics of the flow as described in \citep{Godon96b}. 
Thus outer boundary conditions were imposed directly  
on the incoming characteristics, and computed values from the variables inside 
the domain were imposed (propagated) on the outgoing characteristics.    
The strategy works well, especially when combined with Riemann solvers
which also work on the same principle of resolving the inflowing and outflowing waves. The equatorial boundary condition in the $\theta$-direction
was reflective (symmetric) and the boundary condition at the other 
$\theta$-directional boundary was specified as continue.

\clearpage
\begin{table}
\begin{tabular}{ccc}
\hline
model & $\alpha$ & $\Phi$ \\
\hline
v1 & 0.1   & yes  \\
v2 & 0.03  & yes  \\
v3 & 0.01  & yes  \\
v4 & 0.005 & yes  \\
v5 & 0.001 & yes  \\
nv1 & 0.1   &no    \\
nv2 & 0.03  &no    \\
nv3 & 0.01  &no   \\
nv4 & 0.005 &no   \\
nv5 & 0.001 &no  \\  
\hline
\end{tabular}
\caption{This table shows our ten models. The first column indicates 
the model name. The second column indicates the $\alpha$-viscosity 
employed in the model and the last column shows whether dissipative 
heating was included in the model.}
\end{table}

\clearpage 

The spatially and temporally second order algorithms
in \verb+RIEMANN+ have been described in
\cite{Roe96,Balsara98a,Balsara98b,Balsara99a,Balsara99b,Balsara04}
and use many ideas from higher order WENO schemes
(see \cite{Jiang96,Balsara00}) to reduce dissipation.
The matter in the model is subject to the central gravitational field
of the underlying WD. 
The model uses an ideal gas ($\gamma=5/3$) of a fully ionized
solar composition ($\mu=0.62\,\textrm{g/mole}$) and assumes no
radiative transport.

\section{Results and Discussions}\label{sec:results}
In the next subsections, we focus on several important aspects 
of the boundary layer dynamics as follows.
In section~\ref{subsec:den} we focus on the density profile of the boundary 
layer.
In section~\ref{subsec:prs} we consider the pressure profiles in 
the boundary layer.
In section~\ref{subsec:ang_pol}, we focus mainly on the the evolution of 
angular momentum in the boundary layer .
In section~\ref{subsec:ricnum} 
we check the stability of the flow in the boundary layer. 
In section~\ref{subsec:grav}, we discuss accretion based instabilities 
and in section~\ref{subsec:optical} we relate 
the computations to observations.

\subsection{Density Structure of the Boundary Layer}\label{subsec:den}

Figs 1a and 1b show the logarithm of the density in 
the boundary layer with $\alpha=0.1$ and $\alpha=0.001$ respectively.
Figs 1c and 1d and do the same for the runs 
with the same values of $\alpha$ but with
the viscous heating switched off in eq.~[\ref{eq:heatequation}].
The full extent of the
computational domain is shown in the $\theta$-direction. Only a
small portion of the inner radial direction is shown and the values on
the x-axis of the plot are in units of $10^{9} cm$, making it possible to
measure the radial coordinate . The same convention for labeling figures
applies to all other figures in this paper where flow variables are imaged.

\begin{figure}
\caption{
(a) left panel. 
The logarithmic density for model 
v1 is shown after one Keplerian rotation. 
(b) right panel. 
The logarithmic density for model 
v5 is shown after one Keplerian rotation.  
(c) next page - left panel. 
The logarithmic density for model 
nv1 is shown after one Keplerian rotation.  
(d) next page - right panel. 
The logarithmic density for model 
nv5 is shown after one Keplerian rotation. The full extent of the
computational domain is shown in the $\theta$-direction. Only a
portion of the radial direction is shown and the values on
the x-axis of the plot are in units of $10^{9} cm$ ,
making it possible to measure the radial coordinate. 
} 
\end{figure}

In all cases,
the poloidal velocity is overlaid as vectors, enabling us to trace
the flow of matter on the surface of the star. Thus Fig. 1a 
pertains to the optically thick limit (outburst state) while 
Fig. 1d corresponds to
the optically thin limit (quiescent state). 
We see that a thick, dense boundary layer
forms in both Figs. 1a and 1c while that is not the case in Figs. 1b and 1d.
Thus, physically thick boundary layers form in outburst
and the result is independent of whether the viscous energy 
generation is included in the energy equation. The boundary layers
that form in quiescence tend to be physically thin.
This shows that the viscosity is the primary discriminant in determining
the structure of the BL. The poloidal velocity vectors in Figs 1a
and 1c also show us that the velocity increases with increasing
distance from the star's surface. The velocity vectors in
Figs. 1b and 1d show a similar trend though it is harder to see because
of the smaller physical extent of the BL. Such a velocity structure is also
known to occur in terrestrial fluid dynamics when a viscous fluid
flows over a stationary solid surface, forming a boundary layer at
the solid's surface. This shows us that our simulations are producing
a valid result which is consistent with our intuition. It also shows
us that our decision to refer to these star-disk layers as boundary
layers is well-motivated.

The accretion rate is high enough in the $\alpha=0.1$ cases that the
infalling matter depresses the stellar atmosphere close to the equator.
While this is not so evident in Fig. 1a--1d, we will show in
section~\ref{subsec:grav} that the infalling matter can excite gravity
waves on the surface of the star. Likewise, in section~\ref{subsec:ang_pol}
we will show that the high $\alpha$ case spins up a significant portion
of the star's atmosphere while itself being spun down. The net result
of this process is that the toroidal velocity of 
the disk becomes sub-Keplerian at
larger radii from the star as $\alpha$ increases, resulting in broader boundary
layers. Fig. 2 shows the mass accreted as a function of time.
This figure was generated by integrating the mass from 
the disk-star interface to the top of the BL where $dv_\phi/dr\equiv 0$.
Fig. 2 shows us that the 
accretion rate is higher in the high-$\alpha$ cases, as expected.
Fig. 2 also shows us that for the high $\alpha$ cases the accretion rate
seems to drop off with time. This is entirely a result of the fact that
this round of simulations does not include radiative transfer. As a result,
the base of the boundary layer heats up, with a corresponding increase
in pressure. The radiative cooling times in a real accretion disk are short
enough (i.e. even smaller than an orbital time) that the boundary
layer would cool down quite rapidly. The consequent pressure reduction
would then permit accretion to proceed unimpeded.

\begin{figure}
\caption{The mass accretion is shown as a function of 
time for models v1--v5.} 
\end{figure}

Because of resolution constraints, we had to use temperatures that
are somewhat larger than those that prevail on accreting white dwarfs.
In subsequent work, we intend to overcome this constraint.
It is, nevertheless, interesting to relate physical 
depth to optical depth of the
boundary layer. In doing that, it is important to exclude the disk fluid
that is spinning close to the Keplerian velocity. We, therefore, define
the disk-boundary layer interface as the region where the gradient of the
angular velocity vanishes: $\partial \Omega / \partial r = 0$ at
$r=r_m$.
In the boundary layer ($r<r_m$) the  matter is
spinning with a toroidal velocity that
is smaller than the Keplerian velocity.
For that sub-Keplerian accreted material, we plot
the optical depth due to Thompson scattering as a function of meridional
angle in Fig. 3.
Here we define the optical depth as $-\int_\infty^0 \kappa 
\rho_{disk}dr$, where $\kappa=0.34\,\textrm{g}\,\textrm{cm}^{-3}$ 
is the Thomson scattering opacity of a fully ionized solar composition. 
This constitutes the minimum amount of scattering and thus provides 
a lower bound of the opaqueness of the matter.

The boundary layer is technically defined by the radius where the radial 
gradient of the accretion flow's angular velocity disappears. We use
that definition for the rest of this paper. Because we track the fluids that
make the disk, halo and star, we are in a position to isolate just the
disk material that is within the boundary layer. The optical depth of this
material (in the radial direction) is very important because it sets the
emission characteristics of the accretion belts that have been identified
in the observational literature. Fig. 3 shows us that physically thick boundary
layers in the high-$\alpha$ limit also tend to be optically thick while
physically thin boundary layers in the low-$\alpha$ limit tend to be
optically thin. The physical implication of that is that during outburst
the BL could suppress the emission from a fraction of the star's surface.
If the BL is optically thick with an optical depth $\gg 1$ then the
hottest part of the boundary layer, which prevails at the disk-star interface
would not be visible. However, during both the transitions, from outburst
to quiescence and vice-versa, it is possible that this hot layer might become
visible, giving one a direct view of boundary layer heating.
This might show up as an additional UV component such as the one
seen by \cite{Sion04c} in VW Hydri during its transition from
quiescence to outburst. A similar UV component has been detected in
U Gem by \cite{Long93,Froning01,Szkody02} during its transition from
outburst to quiescence. Obtaining matched measurements of the velocity
and UV excesses would allow one to further corroborate the scenario 
presented here. 
We see that the model in Fig. 1a produces an optically thick
boundary layer during outburst.
It is, therefore, consistent with our claim in Section 2 that inclusion
of the viscous dissipation term in eq.~[\ref{eq:heatequation}] provides
a rather realistic representation of the structure of the 
boundary layer in outburst. Likewise, the model in
Fig. 1d, results in an optically thin boundary layer during quiescence.
It is thus consistent with our claim that excluding the viscous dissipation term in eq.~[\ref{eq:heatequation}] is similarly more representative of
the structure of the boundary layer in quiescence.

 
\begin{figure}
\caption{The optical depth along the radial 
direction is shown for models v1--v5 as a function of 
the angle from the equatorial plane.} 
\end{figure}


\subsection{Pressure in the Boundary Layer}\label{subsec:prs}

Figs 4a and 4b show the logarithm of the pressure in the boundary 
layer with $\alpha=0.1$ and $\alpha=0.001$ respectively.
Figs 4c and 4d do the same for the runs 
with the same values of $\alpha$ but with
the viscous heating switched off in eq.~[\ref{eq:heatequation}].

\begin{figure}
\caption{
(a) left panel. 
This figure shows the logarithmic pressure for model v1 after 
one Keplerian rotation.  
(b) right panel. 
This figure shows the logarithmic pressure for model v5 after 
one Keplerian rotation.  
(c) next page - left panel. 
This figure shows the logarithmic pressure for model nv1 after 
one Keplerian rotation.  
(d) next page - right panel.   
This figure shows the logarithmic pressure for model nv5 after 
one Keplerian rotation.  
} 
\end{figure}




From Figs. 4a and 4b we see that the pressure of the accreted fluid 
is highest in the equatorial
plane. This high pressure can be attributed to a combination of viscous
dissipation as well as the ram pressure due to infall of accreting material.
The pressure gradient in the meridional direction on the surface of the
star, therefore, drives the meridional flow that develops on the WD's surface.
The models that were used in Figs. 4c and 4d did not include viscous heating
in eq.~[\ref{eq:heatequation}]. They, nevertheless, show that the pressure
of the accreted fluid is highest in the equatorial plane and in this
instance, the increased pressure is entirely attributable to the
ram pressure due to infall of accreting material. We also see that the
equatorial pressure increase in Fig. 4b is less than 
that in Fig. 4a and, similarly, for Figs. 4b and 4d. 
The smaller pressures in Figs. 4c and
4d relative to Figs. 4a and 4b can be attributed to the 
additional contribution from viscous heating.
Fig. 4a--4d, therefore, serves
to show us that the formation of the pressure gradient in the meridional
direction, which then drives the meridional flow in the boundary layer,
is a very commonplace phenomenon. In other words, any flow that is put
in a similar geometry and is made to experience similar viscous stresses would
naturally form a similar boundary layer, showing the ubiquity of boundary
layer formation. (The only other requirement that such a BL flow has to satisfy
as the disk material migrates polewards on the star's surface is the ability to
lose angular momentum. The next sub-section will show that it can accomplish this
very efficiently by spinning up the underlying layers of the stellar atmosphere.)
Thus the decision by \cite{Inogamov99,Piro04} to include
a non-zero meridional velocity in their models was of central importance
in forming the inherently multi-dimensional boundary layers that we
see in our simulations. Such boundary layers
also form in accreting neutron stars, and TTauri stars that are
going through the FU Orionis phenomenon
\citet{Kley96,Kley99,Balsara05}. 

\subsection{Toroidal and Poloidal Velocities}\label{subsec:ang_pol}

As the simulations start, the shear force transfers angular momentum 
between differentially rotating disk annuli so that matter can move 
inwards and accrete on the star at the footpoint of the disk.
Angular momentum transfer between the disk material and the stellar 
surface is also necessary so matter can move to higher latitudes. 
Otherwise, the centrifugal barrier prevents matter from leaving the 
footpoint of the disk. It is this transfer of angular momentum which 
spins up the existing surface layers of the star.
Angular momentum is also directly advected onto the star, because the 
orbiting disk material eventually accretes and forms the new surface.

Fig.5a shows the resulting specific angular momentum after 3/4 of 
a Keplerian evolution for the $\alpha=0.1$ case, whereas 
fig.5b shows it for the $\alpha=0.001$ case.
Fig.5a shows that the halo is spun up as is the underlying 
star at the footpoint, where the disk connects with the star. 
Moreover, we observe that by this
time in the simulation the disk material has spun 
up all five atmospheric scale heights
of the star at equatorial latitudes. At higher 
latitudes the spreading disk material
is less dense and thus takes longer amounts of time 
to drag the star's surface around with it.
In fig.5b, which is at the same time as fig.5a and uses
the same color scale, we see that the disk has not spun up 
the star's atmosphere even at the stellar equator.


\begin{figure}
\caption{
(a) left panel.  
A color plot of the specific angular momentum, 
$\Omega=v_\phi/r$, for the $\alpha=0.1$ case after 3/4 of 
a Keplerian rotation period.  
(b) right panel. 
A color plot of the specific angular momentum for 
the $\alpha=0.001$ case after 3/4 of a Keplerian rotation 
period.  
} 
\end{figure}


Fig.6 shows the toroidal velocity and sound speed in 
the disk's midplane as a function of radius for the simulated 
alpha-viscosities.
The cases of $\alpha=0.01$, $\alpha=0.005$, and $\alpha=0.001$ do not 
show any significant spin-up of the star after 3/4 of a Keplerian 
rotation, whence there is not significant shear connection between the
disk and the stellar surface. Therefore disk matter retains its Keplerian 
motion much closer to the star which means that even though the accretion 
rates are smaller, significant amounts of the dissipation can still be
generated in the BL close to the star.

\begin{figure}
\caption{
Toroidal ($\phi$) velocity in the disk's midplane
shows the extent of the boundary layer as well 
as the sound speed after 3/4 of a Keplerian rotation period. 
Note that $v_\phi$ asymptotes to a higher value than 
$c_{s,\alpha}$.} 
\end{figure}

The matter in the layers of the star that we simulate is indeed
non-degenerate. Even so, the high temperatures cause it to have
a rather low specific molecular viscosity. As a result, turbulence and/or
threaded magnetic fields might nevertheless be the two most prominent ways to 
transfer a significant amount of angular momentum to the deeper 
layers of the WD \citep{Durisen73a}, where the relative motion 
drives the mixing between the accreted surface material and the 
deeper layers.
A supersonic component in the toroidal direction over most of the 
BL is obtained for all values of $\alpha$ (see fig.6).
The supersonic toroidal velocities mean 
that the flows are susceptible to the rapid
development of gravity wave and/or Kelvin-Helmholtz 
instabilities in three dimensions.
Such instabilities can be studied by taking a small slice of the interface
from our simulations and extending it in the toroidal direction 
in a three dimensional
simulation. Our present simulations could then provide 
the velocities which drive
such instabilities. The calculation of the turbulent 
mixing which obtains from these
instabilities must therefore be calculated using other
models \citep{Kippenhahn78,MacDonald83,Rosner01,Alexakis04}.

Fig. 7a shows the poloidal Mach number of the accreted material
on the surface of the star for $\alpha=0.1$. Fig. 7b shows the same
for $\alpha=0.001$. The same scale is used for both figures. We see that
the Mach number is mildly transonic in Fig. 7a while it is subsonic
in Fig. 7b.
Only the $\alpha=0.1$ case has a transonic component in the poloidal 
directions, whereas the poloidal flows for $\alpha=0.03$, $\alpha=0.01$, 
$\alpha=0.005$, and $\alpha=0.001$ remain subsonic (see figs. 5a
and 5b) \citep{Fisker05c}.

 
\begin{figure}
\caption{ 
(a) Left panel. 
This figure shows the poloidal Mach number with poloidal 
velocity vectors overlaid for model v1 after one Keplerian rotation. 
(b) Right panel. 
This figure shows the poloidal Mach number with poloidal 
velocity vectors overlaid for model v5 after one Keplerian rotation. }
\end{figure}


\subsection{Hydrodynamic Instability in the Boundary Layer}
\label{subsec:ricnum}

The present simulations are based on an alpha-viscosity formulation,
which implicitly assumes an underlying model for the sub-scale turbulence.
However, the material that accretes on to a white dwarf has very
low viscosity. Observations have not revealed the existence
of magnetic fields at the WD surface. Besides, the radial gradient
of the angular velocity has the wrong sign for the MRI to act.
It is, therefore, interesting to explore the role of other hydrodynamical
instabilities at the surface of the star. It is important to remember
that the alpha-viscosity formulation does smear the velocity gradients.
Yet, if these gradients are amenable to the development of a persistent
hydrodynamical instability then the simulation should reveal its existence
to us. It is in that spirit that we try to identify an instability
that might give rise to a sub-scale hydrodynamical turbulence on
an accreting white dwarf's surface.

There are several instabilities 
that can possibly lead to turbulence in boundary layers, such as 
the centrifugal instability (Rayleigh's criterion),
the shearing Kelvin-Helmholtz instability, or even the 
Rayleigh-Taylor instability (buoyancy forces). Depending on the 
exact angular velocity profile, pressure and density gradients 
each of the forces involved compete, some are stabilizing 
while other are destabilizing.  
The sufficient condition for linear stability of 
a rotating, radially stratified fluid under 
the influence of (a radial) gravity was given by \citet{Sung1974}      
\begin{equation} 
\frac{1}{\rho} g_{eff} S_{\varpi} + 
\frac{k_z^2}{M} \Phi - \frac{1}{4} \frac{m^2}{M} 
\left( \frac{d \Omega}{d \varpi } \right)^2 \ge 0 , 
\end{equation} 
where, $m$ and $k_z$ are the angular and vertical mode numbers
(in cylindrical coordinates $(\varpi, \phi, z)$), 
$M$ is defined as $M= k_z^2 + m^2/\varpi^2$,  
$g_{eff}$ is the effective gravity defined as  
$$
g_{eff} = \frac{GM}{\varpi^2}- \varpi \Omega^2 ;
$$ 
$S_{\varpi}$ is the Schwarzschild discriminant \citep{Schwarzschild1906}
$$
S_{\varpi} = \left( \frac{d \rho}{d\varpi} \right)_{ad}
- \left( \frac{d \rho}{d\varpi} \right) , 
$$
in which 
$$
\left( \frac{ d \rho}{d\varpi} \right)_{ad} = \frac{\rho}{\Gamma_1 p} 
\frac{dp}{d\varpi} 
$$ 
is the adiabatic density gradient;
and $\Phi$ is the Rayleigh discriminant \citep{Rayleigh1880,Rayleigh16}
$$
\Phi = \frac{1}{\varpi^3} \frac{d}{d\varpi} \left( 
\varpi^2 \Omega \right)^2 . 
$$ 
In the present case we do not consider vertical modes $k_z$ and
only consider the equatorial flow for which $r=\varpi$. 
There are two distinct cases as follows.  

The axi-symmetric
case (mode $m=0$) gives the Solberg-H\o iland criterion 
\citep{Sung1974,Sung1975} 
$$
\frac{1}{\rho} g_{eff} S_{r} + \Phi \ge 0. 
$$
The perturbations of an accretion disk to a convective
instability has already been studied in this context
\citep{Livio1977,Rudiger2002,Johnson2006}.
In the limit of vanishing rotational velocity, the Solberg-H\o iland
criterion simply leads to the classical Schwarzschild condition
$S_r > 0$
\citep{Lebovitz1965,Lebovitz1966}, which is also the condition 
in the vertical direction in the disk \citep{Livio1977}.
In the limit of a vanishing gradient of the density,
the Solberg-H\o iland criterion 
simplifies to Rayleigh's criterion for rotating fluids.  
The axi-symmetric instability in the boundary layer 
could possibly lead to a ring-like
structure oscillating and in many ways similar to the ``breathing'' 
mode of an unstable star. This mode will not lead to turbulence,
but could it explain the short period oscillations observed in CVs?  
The mode would depend  on the density, temperature and rotation rate 
(Solberg-H\o iland criterion), and  
its period and coherence would vary with these parameters.  
However, dwarf nova oscillations (DNOs) in CVs 
exhibit a 180deg phase jump at eclipse \citep{pat79},  
and frequency doubling (presence of the first harmonic
\citet{pat81}, and therefore they 
cannot be explained by an $m=0$ mode.  
On the other hand, Quasi-periodic oscillations (QPOs) in CVs  
do not all exhibit the same
characteristics (some are believed to form at larger radii in the
disk), and, therefore, 
we cannot completely rule out the $m=0$ mode 
in the boundary layer as an additional
mechanism to produce quasi-periodic oscillations (QPOs). 

We now turn our attention to the non-axisymmetric case.  
In the case $m\ne 0$, a sufficient condition for stability that does
not involve $k_z$ is \citep{Sung1974}
\begin{equation} 
\frac{1}{\rho} g_{eff} S_r -\frac{1}{4} r^2 
\left( \frac{d \Omega}{dr} \right)^2 \ge 0. 
\end{equation}
This condition is usually written in the form of a modified
Richardson number 
\begin{equation} 
R_i = 
g_{eff} 
\left( 
\frac{1}{\Gamma_1 p} \frac{dp}{dr} - \frac{1}{\rho}\frac{d \rho}{dr} 
\right) 
\left( r \frac{d \Omega}{dr} \right)^{-2}  
 \ge  \frac{1}{4}.  
\end{equation}
This is a generalization of the Miles-Howard theorem
(\citet{Miles1961,Howard1961}; see also \citet{Chimonas1970}). 
The Miles-Howard theorem itself reduces to the original Richardson
criterion \citep{Richardson1920}  
when compressibility is omitted and the Schwarzschild 
discriminant is replaced by the buoyancy term alone.  
The modified Richardson number was considered in a few analytical
studies to analyze the stability of material accreting on the surface
of a white dwarf \citep{Durisen1977,Kippenhahn78,Livio1987}.  
Here, we have the opportunity to use results from numerical simulations
to evaluate the modified Richardson number in the boundary layer.  
Using the definition of the Ledoux discriminant \citep{Ledoux1958}, 
$$ 
A = \frac{1}{\Gamma_1} \frac{d~ln~P}{dr} - \frac{d~ln~\rho}{dr}, 
$$ 
the (modified) Richardson condition for stability is then written 
\begin{equation}
R_i = N^2_{BV}  \left( r \frac{ d\Omega}{dr} \right)^{-2}  \ge 
\frac{1}{4} , 
\end{equation}  
where $A$ is related to the buoyancy (or Brunt-V\"ais\"al\"a) 
frequency $N_{BV}$
\citep[e.g.]{Pesnell1986,Livio1987}
$$
N_{BV}^2 = g_{eff}A. 
$$ 
The the flow is unstable for  $R_i < 1/4$.  
In Fig.8 we show the Richardson number, 
$R_i$ in the $\phi$-direction (eqs. 10, 11), for run 91 and 95.
The Richardson number is smaller than 1/4 in most of the domain 
and therefore the flow is unstable. It is important to remark 
that while the shear could stabilize the flow \citep{Johnson2006}, 
internal gravity waves can be reflected from a shear layer \citep{Van1982}, 
and can be over-reflected from a rigid boundary \citep{Sachdev1982}.   
It has also been shown that modes can be unstable at the star-disk
interface due to the propagation through the corotation \citep{tsa09}.  
Unstable modes could be trapped and grow between (i) the 
surface of the WD and the strong shear region; and/or (ii) between the 
strong shear region and the corotation radius at larger radii in the
inner disk, in a manner similar to the Papaloizou-Pringle instability
\citep{papa84,papa85,papa87} observed in simulations of disks with a
rigid inner boundary \citep{Godon1997}. 
The question of how the instability
will develop cannot be addressed without full 3D simulations, which 
are beyond the scope of this work. 
However, a Richardson number $<1/4$
in the boundary layer region and {\it at} the stellar surface raises
the possibility of an instability. This instability
could  develop in the form of waves in the ``spread layer'' as studied by 
\citet{pir04b} to explain DNOs in CVs, and raises the possibility of     
strong mixing between the hydrogen rich freshly
accreted material and the WD material.


\begin{figure}
\caption{
The toroidal Richardson Number $R_{\phi}$ as a function of the
radius $r$ in the equatorial plane ($\theta = 90^{\circ}$), for
run 91 and 95. The stellar radius is located at $r=9 \times 10^8$cm. 
In most of the domain the Richardson number is smaller than 1/4.  
} 
\end{figure}


\subsection{Formation of surface gravity waves}\label{subsec:grav}

For the $\alpha=0.1$ case, the outer edge of the boundary layer 
(where $d\Omega/dr\sim 0$) is located at $1.06r_*$. 
At $r=1.03r_*$, $v_\phi$ is about 92\% of $v_K$. This means that 
only 84\% of the gravitational pull is supported by the centrifugal 
force while the rest is supported by pressure.
The pressure for $\alpha=0.1$ is illustrated in fig.4a and comes from the 
build up of density due to the high accretion rate facilitated by the high 
viscosity which drags down material from the disk and also keeps it 
from moving rapidly towards the poles once it makes contact with the WD 
surface.
This results in a dense band at the foot point of the disk which 
causes a surface gravity wave of surface matter to spread towards the poles. 
Such surface gravity waves have been studied in the context of terrestrial
physics by \cite{Miles57} and the role of turbulence in exciting these
waves has also been studied \cite{Phillips57} and the simulations
show a similar phenomenon on the surface of strongly accreting
white dwarfs.
The matter inflow is, however, sufficiently large to cause the disk 
material to overflow the gravity wave supersonically as described 
above and shown in fig.5a. Our present simulations do not show any
evidence of wave breaking. However, should wave breaking reveal itself
in simulations where the accretion is more vigorous, it would provide
a further mechanism for directly mixing accreted material with the
CNO-rich material that lies beneath the surface of the white dwarf.

The propensity for the creation of gravity waves and their magnitude 
decreases rapidly with lower values of $\alpha$. Even for $\alpha=0.03$, 
the effect is only marginal and for lower values it is no longer present. 
The effect is thus only present for large values of $\alpha$, and 
possibly during dwarf novae in the outburst stage. 
Furthermore the gravity wave might be transient as the surface adjusts to 
a fluctuations in the accretion rate. For lower values of $\alpha$, the 
matter accretes slowly inwards and spreads towards the poles in a uniformly 
thick layer.

\subsection{Dissipative heating of the BL}\label{subsec:optical}

The local rate of energy dissipation depends on the divergence of the angular velocity \citep{Mihalas84}, so fig.6 also indirectly shows where heat 
is released.
Fig.9a shows that the supersonic impact of the accretion flow dissipates 
the most energy. However, energy is also dissipated at the interface 
between the BL and the surface. This is better seen in 
Fig.9b which shows that the toroidally rotating 
matter dissipates rotational kinetic energy along its entire 
interface with the surface as it slides against it.

In situations where the BL is optically thin, the radiation
will be emitted from the very same regions where the heat is generated
in fig. 6b. This may explain 
the optically thin BL and line emission seen in observations of 
U Gem during the quiescent phase (see also: \citet{Fisker05c}).
Even in outburst, Fig. 3 shows that the optically thick boundary layer
has an optical depth of ~ 100. We therefore expect that the radiative 
diffusion of the heat in the radial direction will be faster than its 
advection in the poloidal direction. Consequently, our simulations 
provide a dynamically consistent reason for the formation of a warm 
belt of accreted matter, as was suggested in the observations of 
\citet{Sion04c} and the modeling of \citet{Godon05}.


\begin{figure}
\caption{
(a) Left panel. 
A logarithmic color plot of the dissipation rate for 
the $\alpha=0.1$ case after 3/4 of a Keplerian rotation period. 
(b) Right panel. 
A logarithmic color plot of the dissipation rate for the 
$\alpha=0.001$ case after 3/4 of a Keplerian rotation period. 
Notice that the color scale is different for fig. 8a.   
} 
\end{figure}


\section{Summary and Conclusion}\label{sec:conclusions}

We have numerically simulated the structural dynamics of the BL for an
accreting white dwarf surrounded by an $\alpha$-disk for different
values of $\alpha$.  The structure and dynamics of the BL is important,
because it determines the specifics of the radiated spectrum that is
emitted from this region. The energy radiated from the
BL could account for up to half of the total energy released in the
accretion process.
For high values of $\alpha$, the BL is optically thick and extends more
than 30 degrees to either side of the disk plane after 3/4 of a
Keplerian rotation period ($t_K$=19s).  

The simulations also show that high values of $\alpha$ result in a
spreading BL which sets off gravity waves in the surface matter.
The accretion rate can be high enough
in the high $\alpha$ cases to cause a depression to form on the surface
of the star. The accretion flow moves supersonically
over the depression that is formed, making it susceptible to the rapid
development of gravity wave and/or Kelvin-Helmholtz instabilities.  
The low viscosity cases also show a spreading BL, but here the accretion
flow does not set off gravity waves and it is optically thin.

We argued in the paper by \citet{Sion04c} 
that an accretion belt might be sustained after a
long period of perhaps thousands of dwarf nova events such that
an equilibrium or steady state is established between the Richardson
number and the average rate of accretion.  Our 2D simulations in
this paper has not been followed long enough for any steady state
behavior to be observed.  Therefore, it may be premature for us
on the basis of the limited extent of these first simulations to
argue in favor or against the formation of an accretion belt. 

If the boundary
layer remains optically thick following an episode of high accretion,
this could explain the "second components" of FUV flux seen in several
dwarf novae during quiescence. It should also be pointed out that
a hot equatorial accretion region could persist even after the
material has reached co-rotation with the white dwarf.  For example,
a hot (50,000K) region of low rotation is seen in successive HUT
spectra of U Gem taken early after an outburst and very late after
the same outburst.

The susceptibility of the flow in the boundary layer to undergo a
purely hydrodynamic  instability (modified Richardson number $<1/4$) 
was found assuming an {\it a priori} 
alpha viscosity parametrization consistent with that of the disk. 
The unstable flow would
be very effective in mixing of the accreted material with the outer 
layer (few scale heights) of the WD envelope. 
However, it has been shown \citep{papa94,nara94,kato94,god95} 
that the turbulent viscosity law for
a non-Keplerian disk cannot be represented by the standard alpha
viscosity prescription ($\alpha=$constant), as alpha is proportional
to the shear \citep{god95,abra96} 
$$
\alpha \propto \partial \Omega / \partial r. 
$$ 
Even more so, in the boundary layer, 
the turbulent viscosity is proportional to the 
square of the turbulent Mach number ${\cal{M}}_t^2$ 
\citep{shak88,god95}, which decreases quickly for for subsonic turbulence 
(supersonic turbulence quickly dissipated due to shocks)
$$
\nu_{BL} = \alpha_{BL} c_s H, ~~~~~~~~
\alpha_{BL}= \alpha {\cal{M}}_t^2 \frac{\delta_{BL}}{H} 
\frac{c_s}{v_K-v_*} .  
$$
Supersonic radial infall in the boundary layer also leads to a smaller
turbulent viscosity as shown by \citep{papa94,nara94,kato94,god95},
and the formulation of such a viscosity can also be obtained 
considering causality in one-dimensional steady-state models
of accretion disk boundary layers \citep{poph92}, namely the 
radial infall viscosity cannot be larger than the maximum speed
of the turbulence.
In that case one has \citep{god95}  
$$
\nu_{BL} = \alpha_{BL} c_s H, ~~~~~~~~~~~
\alpha_{BL} = \alpha \frac{\delta_{BL}}{H} 
\frac{1}{{\cal{M}}^3_r},
$$
where ${\cal{M}}_r > 1$ is the radial infall Mach number, and
it is smaller than the turbulent Mach number ${\cal{M}}_t$ 
(see also \citet{poph92}). 

In these regions close to the 
stellar surface (and even more so away from the equatorial region), 
one can use a viscosity of the form
\begin{equation} 
\nu  \propto (v-v_{*}) z , 
\end{equation}   
where $v_*$ is the stellar rotational velocity, $v$ is the velocity
of the flow, 
and $z$ is the distance from the surface of the WD.  It is clear
that such a viscosity would decrease rapidly 
as $z \rightarrow 0$, and would agree with the fact that the size 
of the turbulent Eddies be limited by the presence of the stellar
surface and their velocity proportional to the change of $v$ over 
a distance $z$ (see also \citet{Landau87}).

Piro and Bildsten have shown that the properties of the boundary
(``spread'') layer vary depending on the value of the viscosity,
however, like us, they assume an alpha viscosity in which alpha is not a
function of the coordinates, the radial infall velocity or
or the shear. 
It is our aim, in future simulations, to carry out simulation of
the boundary layer, by assuming
a viscosity law that is greatly reduced in the boundary layer. 
For this purpose, 
one could use a combination of viscosity prescriptions suggested
in the above mentioned works. Since the results presented here
as well as Piro \& Bildsten's results depend strongly on the value
of $\alpha$ (which was assumed to be constant), it is clear that 
a modified alpha viscosity prescription can lead to some new
and unexpected results.

The evolutionary simulations presented here, while followed for only a
brief interval of time, represent the first successful, fully hydrodynamic  
treatment of the the flow of accreted matter {\it into}
the white dwarf surface
layers. All previous attempts to follow accretion hydrodynamically from
the boundary layer into the white dwarf surface layers were either not
computationally viable or the underlying white dwarf was treated as a
solid boundary.

The successful convergence of our dynamical model simulations opens the
door to going well beyond this first stage and ultimately follow the
dynamical evolution over much longer time intervals for both the high
viscosity and low viscosity cases with the inclusion of radiative processes.

\acknowledgements
PG wishes to thank Steve Lubow for a discussion on the importance of the 
adiabatic term in the modified Richardson number, and 
Mario Livio for his kind hospitality at 
the Space Telescope Science Institute.
This work is supported by NASA ATFP grant NNX08AG69G to Villanova University
and the University of Notre Dame.  
Participation by EMS and PG was also supported in part by NSF grant
AST08-07892 and NASA ADP grant NNX04GE78G to villanova University.

\appendix
\section{Numerical model}\label{appendix:model}
The source of the angular momentum transport in the BL is a combination of magnetic fields and turbulence. However, an a priori prescription, in particular in the BL, is a source of disagreement (see \cite{Popham95} and references therein). Instead the efficiency of the angular momentum transport can be parametrized with a coefficient, $\alpha$ \citep{Shakura73}.
Describing the angular momentum transport with a simple shear coefficient 
means that the dynamics follows the Navier-Stokes equations.
Here the Navier-Stokes equations are solved in spherical coordinates $i=(r,\theta,\phi)$ as given by \cite{Mihalas84}. The accelerations are defined by the time-derivative of the velocities which are given in spherical coordinates.
\begin{eqnarray}
\rho a_r & = & -\frac{GM\rho}{r}-\frac{\partial p}{\partial r}+\frac{\partial }{\partial r}\left[2\mu \frac{\partial v_r}{\partial r}+\left(\zeta-{2 \over 3}\mu\right)(\nabla\cdot v)\right] + {1\over r}\frac{\partial }{\partial \theta}\left\{\mu\left[r\frac{\partial}{\partial r}\left({v_\theta \over r}\right)+{1 \over r}\frac{\partial v_r}{\partial \theta}\right]\right\} \nonumber \\
&+&{1 \over r \sin \theta} \frac{\partial }{\partial \phi} \left\{\mu\left[{1 \over r \sin \theta}\frac{\partial v_r}{\partial \phi} + r \frac{\partial }{\partial r}\left({v \over v_\phi}\right)\right]\right\}\nonumber \\
&+&{\mu \over r}\left[4r\frac{\partial }{\partial r}\left({v_r \over r}\right)-{2\over r \sin \theta}\frac{\partial }{\partial \theta} (v_\theta \sin \theta ) {2 \over r \sin \theta} \frac{\partial v_\phi}{\partial \phi}+r\cot\theta\frac{\partial }{\partial r}\left({v_\theta \over r}\right)+{\cot \theta \over r}\frac{\partial v_r}{\partial \theta}\right]
\end{eqnarray}
\begin{eqnarray}
\rho a_\theta &=& -{1 \over r}\frac{\partial p}{\partial \theta}+\frac{\partial }{\partial r}\left\{\mu\left[r\frac{\partial }{\partial r}\left({v_\theta \over r}\right)+{1 \over r}\frac{\partial v_r}{\partial \theta}\right]\right\}+{1 \over r}\frac{\partial }{\partial \theta}\left[{2\mu \over r}\left(\frac{\partial v_\theta}{\partial \theta}+v_r\right)+\left(\zeta-{2 \over 3}\mu\right)(\nabla\cdot v)\right]\nonumber \\
&+&{1 \over r \sin\theta}\frac{\partial }{\partial \phi}\left\{\mu\left[{\sin \theta \over r}\frac{\partial }{\partial \theta} \left({v_\theta \over \sin \theta}\right)+{1 \over r \sin \theta}\frac{\partial v_\phi}{\partial \phi}\right]\right\}+{\mu \over r}\left\{{2 \cot \theta \over r}\left[\sin \theta \frac{\partial }{\partial \theta} \left({v_\theta \over \sin \theta}\right)-{1 \over \sin\theta}\frac{\partial v_\phi}{\partial \phi}\right]\right.\nonumber \\
&+&\left.3r\frac{\partial }{\partial r}\left({v_\theta\over r}\right)+{3 \over r}\frac{\partial v_r}{\partial \theta}\right\}
\end{eqnarray}
\begin{eqnarray}
\rho a_\phi &=& - {1 \over r \sin \theta} \frac{\partial p}{\partial \phi}+\frac{\partial }{\partial r}\left\{\mu\left[{1 \over r \sin \theta}\frac{\partial v_r}{\partial \phi}+r\frac{\partial }{\partial r}\left({v_\phi \over r}\right)\right]\right\}+{1 \over r}\frac{\partial }{\partial \theta}\left\{\mu\left[{\sin \theta \over r}\frac{\partial }{\partial \theta}\left({v_\phi \over \sin \theta}\right)+{1 \over r\sin\theta}\frac{\partial v_\theta}{\partial \phi}\right]\right\}\nonumber \\
&+&{1 \over r\sin\theta}\frac{\partial }{\partial \phi}\left[{2\mu\over r}\left({1 \over \sin\theta}\frac{\partial v_\phi}{\partial \phi}+v_R+V_\theta\cot\theta\right)+\left(\zeta-{2 \over 3}\mu\right)(\nabla\cdot v)\right]\nonumber \\
&+&{\mu \over r}\left\{{3 \over r \sin \theta}\frac{\partial v_r}{\partial \phi}+3r\frac{\partial }{\partial r}\left({v_\phi \over r}\right)+2\cot\theta\left[{\sin\theta\over r}\frac{\partial }{\partial \theta}\left({v_\theta\over\sin\theta}\right)+{1 \over r \sin \theta}\frac{\partial v_\theta}{\partial \phi}\right]\right\}
\end{eqnarray}
Here $a_i$ are the accelations, $p$ is the pressure, and $\zeta$ is the coefficient of bulk viscosity, which we set to zero to treat the fluid as a Maxwellian fluid. The kinematic viscosity $\nu=\mu/\rho$ is set by the alpha-disk prescription of \cite{Shakura73}, so $\nu=\alpha c_sH$, where $H$ is the disk scale height which should properly correspond to the turbulent turnover scale with convective bubbles moving at sound speed $c_s$.

The viscous dissipation function is given by
\begin{eqnarray}
\Phi & = & 2\mu\left\{\left(\frac{\partial v_r}{\partial r}\right)^2+\left({1 \over r}\frac{\partial v_\theta}{\partial \theta}+{v_r \over r}\right)^2\right.+\left({1 \over r\sin\theta}\frac{\partial v_\phi}{\partial \phi} +{v_r\over r}+{v_\theta\cot\theta\over r}\right)^2\nonumber\\
&+&\left.{1\over 2}\left[r\frac{\partial }{\partial r}\left({v_\theta \over r}\right)+{1\over r}\frac{\partial v_r}{\partial \theta}\right]^2+{1\over 2}\left[r\frac{\partial }{\partial \phi}\left({v_\phi \over r}\right)+{1\over r\sin\theta}\frac{\partial v_r}{\partial \phi}\right]^2+{1\over 2}\left[{\sin\theta\over r}\frac{\partial }{\partial \theta}\left({v_\phi \over \sin\theta}\right)+{1\over r\sin\theta}\frac{\partial v_\theta}{\partial \phi}\right]^2\right\}\nonumber\\
&+&\left(\zeta-{2 \over 3}\mu\right)\left[{1\over r^2}\frac{\partial }{\partial r}(r^2v_r)+{1\over r\sin\theta}\frac{\partial }{\partial \theta}(v_\theta\sin\theta)+{1\over r\sin\theta}\frac{\partial v_\phi}{\partial \phi}\right]^2\,.
\end{eqnarray}
The dissipation function enters the heat equation as
\begin{equation}\label{eq:heatequation}
\rho\left[c_v\frac{dT}{dt}+p{d\over dt}\left({1 \over \rho}\right) - {GM \over r} \right] = \Phi -\nabla\cdot \vec{q}
\end{equation}
where $T$ is the temperature, $c_v=R/(\gamma-1)$ is the specific heat capacity. As mentioned in section 2, the heat equations does not contain transport terms, so $\vec{q}\equiv 0$. This means that dissipated heat can only be lost be pressure work. Conversely, if the dissipation term is neglected as is the case in the optically thin limit, the only way to change the temperature at any given point is through pressure work.
\section{Computational domain setup}\label{appendix:domainsetup}
We set up an accretion disk around the WD. This setup comprises three parts: the disk, the stellar atmosphere and the halo. This improves on the previous simulations of \cite{Kley87a} which did not include a stellar atmosphere. For stability reasons it is important that the pressure of the three components match at their respective interfaces. This is to prevent shocks from dominating the solution. We have therefore carefully designed the domain to ensure the partial pressures are identical at the common interface.

The radius of the star is $r=9000\textrm{km}$. We assume that the disk is symmetric in the plane and that is it axisymmetric. We use the model of \cite{Balsara04}. In their notation, we have $a=1/2$ and $\gamma=5/3$, so their model effectively reduces to a constant density in the disk 
plane for our case. We assume the the initial halo, atmosphere and disk are isothermal and obey the polytropic equation of state
\begin{equation}\label{eq:polytrope}
P=K\rho^\gamma
\end{equation}
We set $\rho_0=1.2\times10^{-5} \textrm{g}\textrm{cm}^{-3}$. Setting a temperature for the disk then fully determines $K$. Above the disk
plane the partial pressure of the disk matter follows the prescription of a plane parallel atmosphere where the pressure falls off with the e-folding distance given by the scale height. The density is calculated according to eq.~[\ref{eq:polytrope}] 
(see \cite{Landau87}.

The temperature of the WD is set to $T_*=300000\textrm{K}$. The atmospheric scale height is given by $H=RT_*/g\mu$, where $g=GM/r^2$, $R$ is the gas
constant and $\mu$ is the mean molecular weight. The inner boundary of the computational domain is set to $r_*-5H_*$. The disk scale height is given by $H_d=c_s\sqrt{r/g}$, where $c_s=\sqrt{\gamma P/\rho}$ is the local sound speed. The outer boundary of the computational domain is set to $r_*+3H_d$. The disk is set up with a Keplerian velocity profile.
The disk temperature is set to $T_d=1000000\textrm{K}$. Pressure matching between the disk and the stellar atmosphere is achieved by selecting $\rho_*$ so that $P_*\equiv P_d$ at the base of the disk at $r=r_*$. The pressure profile of the WD atmosphere follows the standard solution of a plane parallel atmosphere (see \cite{Landau87})

In these models we do not have a self-consistent disk corona,
which would provide pressure balance at the upper boundary of the accretion
disk. For that reason, we use a very hot, initially static halo that provides
pressure balance both to the upper boundary of the disk as well as to the
WD's atmosphere. The halo's temperature is chosen to be high enough that the
mass in the halo is less than the mass in the disk or the WD atmosphere
by a factor of 35, making it dynamically unimportant.
The temperature of the halo is set so $T_h=50T_d$. The scale height of the halo is determined by the gravitational component in the z-direction. The density of the halo is set to match the pressure of the disk one disk scale height above the disk base at $r_*$.

The careful pressure matching described above ensures that the dynamical instabilities at $t=0$ are negligible. Therefore, the computational model will quickly stabilize to a dynamical equilibrium.

\end{document}